\documentclass[DIV=15,11pt,abstract]{scrartcl}
\usepackage[utf8]{inputenc}
\usepackage{amssymb, mathtools, bm, amsthm, thmtools, relsize}
\usepackage{float}
\usepackage{graphicx} 
\usepackage{authblk}
\usepackage[bottom]{footmisc}   

\usepackage[toc,page]{appendix}
\usepackage[noadjust]{cite} 
\usepackage{xcolor}
\definecolor{myBlue}{RGB}{1,1,141}

\usepackage[colorlinks=True]{hyperref}
\hypersetup{allcolors=myBlue}

\newcommand{\del}{\partial}

\newcommand{\cC}{\mathcal{C}}

\newcommand{\cG}{\mathcal{G}}

\newcommand{\cO}{\mathcal{O}}

\newcommand{\cZ}{\mathcal{Z}}

\newcommand{\ba}{\mathbf{a}}
\newcommand{\bb}{\mathbf{b}}
\newcommand{\bc}{\mathbf{c}}
\newcommand{\bd}{\mathbf{d}}
\newcommand{\be}{\mathbf{e}}
\newcommand{\bf}{\mathbf{f}}

\newcommand{\mus}{\mu^{6\varepsilon}}
\newcommand{\muf}{\mu^{4\varepsilon}}

\newcommand{\dlog}{\mu\partial_\mu}

\title{Renormalization group flow of $O(N)^3$-invariant general sextic tensor model}

\author[1]{G.~Bardy}
\author[2]{T.~Krajewski} 
\author[3]{T.~Muller}
\author[1,4]{A.~Tanasa}

\affil[1]{\textit{Univ. Bordeaux, LaBRI CNRS UMR 5800, Talence, France}}
\affil[2]{\textit{Aix Marseille Univ, Université de Toulon, CNRS, CPT, Marseille, France}}
\affil[3]{\textit{Univ. Sorbonne Paris Nord, LIPN UMR CNRS 7030, Villetaneuse, France}}
\affil[4]{\textit{DFT, H. Hulubei Nat. Inst. Phys. Nucl. Engineering, Magurele,  Magurele, Romania}}

\begin{document}


\maketitle

\begin{abstract}
We compute the beta functions for the $O(N)^3$-invariant general sextic tensor model up to cubic order in the coupling constant, and at leading order in the $1/N$ expansion. 
The method we use is to
identify the appropriate 
Feynman graphs, compute their amplitudes, which then allows us to obtain the $\beta$ functions of the model.
We perform these computations considering both a long-range and a short-range propagator, within the dimensional regularization framework. 
We find three fixed points in the short-range case and a line of fixed points, parameterized by the wheel interaction, in the long-range case. This line of fixed points is 
identical to 
the one found
in the case of the $U(N)^3$-invariant model. Our result shows that
the additional $O(N)^3$-invariant interactions do not modify the long-range fixed point structure of the model.
\end{abstract}


\section{Introduction}Tensor models (see the books \cite{book_gurau, book_tanasa}, or the reviews \cite{Gurau_2012, klebanov2018tasi}) are zero-dimensional quantum field theories where the fundamental fields are rank-$r$ tensors $T_{a_1\cdots a_r}$. They were originally introduced in \cite{sasakura_tensor_1991, ambjorn_three-dimensional_1991, GROSS1992144, BOULATOV_1992} with the aim of generalizing the success of matrix models (see \cite{HOOFT1974461, ginsparg1991matrix, ginsparg1993lectures2dgravity2d, difrancesco20042dquantumgravitymatrix}) to higher dimensions. 

A key tool used to study both matrix and tensor models is the celebrated $1/N$ expansion (see \cite{Moshe_2003}), which organizes the partition function and correlation functions as power series in $1/N$, where $N$ denotes the size of the matrix or resp. the tensor. This expansion technique, also employed in the study of vector models \cite{Eyal_1996}, has proven very useful in the analysis of critical phenomena.

In the matrix model case, the expansion is controlled by the genus. Thus, the matrix Feynman diagram expansion in ribbon graphs corresponds to a sum over two-dimensional discretized surfaces (see again \cite{HOOFT1974461}). The dominant graphs in this expansion have genus zero and are identified with planar surfaces (or combinatorial maps).

The perturbative expansion of tensor models is controlled by the degree \cite{Gurau_2011, gurau_complete_2012}. Dominant graphs in the large $N$ limit have vanishing degree, defining the so-called melonic limit. 

There exists a wide variety of tensor models with different symmetry groups and interaction structures. The action can be constructed to be invariant under the action of the symmetry groups $U(N)$ \cite{Bonzom_2012}, $O(N)$ \cite{Carrozza_2016}, or $Sp(N)$ \cite{Carrozza:2018psc}. The analysis of their large $N$ limits, as well as the implementation of double-scaling mechanisms, has been extensively studied (see for example \cite{  ferrari2019newlargenexpansion, Prakash_2020, Bonzom_2022, Krajewski_2023}). 

This melonic large $N$ limit emerges naturally in the diagrammatic expansion of another type of model, namely the Sachdev-Ye-Kitaev (SYK) model (see \cite{Sachdev_1993} for the original Sachdev-Ye model, the talk \cite{Kitaev_2015} for the holographic proposal and, for example \cite{Rosenhaus_2019, Trunin_2021} for reviews). Remarkably, a reformulation of the SYK model without quenched disorder is possible using a one-dimensional tensor field, thus establishing a direct connection between strongly correlated systems, holography and tensor models \cite{witten2016syklikemodeldisorder, Klebanov_2017, Gurau_2017, Bonzom_2019, GURAU2017386, Krishnan_2017, de_Mello_Koch_2020}.

This connection between tensor model and holography has motivated interest in higher-dimensional generalizations of the original zero-dimensional tensor models. Tensor field theories (TFTs) are thus quantum field theories in which the fundamental fields are rank-$r$ tensors propagating in $d$-dimensional spacetime \cite{Giombi_2017, gurau2019notes}. This framework enables standard field-theoretical  techniques to be applied to tensor models. In particular, the study of renormalization group (RG) flows in TFTs led to the discovery of a non-trivial fixed points. In the large $N$ limit, these fixed points provide candidates for analytically tractable conformal field theories (CFTs) \cite{Bulycheva_2018, benedetti2020melonic}.

On the other hand, 
long-range models \cite{Psak} have proven to be particularly valuable candidate in the search for fixed points of QFT models. Originally introduced in the study of phase transitions (see \cite{Fisher, PhysRevB.15.4344}), they were generalized to vector models \cite{Chai_2021, Slade_2017, Giombi_2023} and, more recently, to tensor field theories \cite{Benedetti_2019, Benedetti_2020, Harribey_2022}. In the context of TFTs, long-range propagation has led to new classes of fixed points that differ qualitatively from the Wilson-Fisher fixed point: instead of isolated fixed points, one finds continuous lines of fixed points parametrized by a single interaction coupling (see again \cite{Benedetti_2019, Benedetti_2020}).

Sextic tensor models have attracted an increasing amount of interest, see for example \cite{Giombi_2018}. In this paper, the large $N$ limit was implemented and RG computations 
were performed. The double-scaling limit was subsequently implemented in \cite{Krajewski_2023}. This 
$O(N)^3$-invariant  
model has a dominant prismatic interaction and is therefore referred to as the prismatic model. 

As shown in \cite{Carrozza_2016}, one can rescale the tensor interactions such that all these interactions contribute at the same order in the large $N$ expansion. The proposed scalings are thus called optimal scalings. The $U(N)^3$-invariant sextic model with such optimal scalings has been studied in both the short-range and long-range regimes in \cite{Benedetti_2020}. At leading order in $1/N$, non-trivial fixed points were identified, including a continuous line of fixed points in the long-range case.

The complete large $N$ limit of the $O(N)^3$ sextic model with such optimal scalings was studied recently in \cite{largeN}.

The same $O(N)^3-$invariant general sextic model 
was analyzed perturbatively in \cite{jepsen2023rg}. The authors identified several short-range fixed points, recovering the $U(N)^3$ fixed points when the additional interactions were turned off. Furthermore, a supplementary prismatic fixed point was discovered. These results were then confirmed in \cite{Fraser-Taliente:2024rql}, where a sextic $O(N)^3$ model coupled to a Yukawa interaction was investigated in detail.\footnote{Note that these papers use different 
approach, see \cite{Osborn:2017ucf, Benedetti:2020rrq}
of the one of our paper.}. 

More precisely, in Section $3$ of \cite{Fraser-Taliente:2024rql}, the authors have performed a standard ${\Bar{MS}}$ calculation of the $\beta$ functions of the sextic $O(N)^3$ model coupled to a Yukawa interaction in $D=3-\varepsilon$ to third order in the coupling constants, both in full generality and specialising to the large $N$ limit.


Let us also emphasize that, despite these extensive studies of this $O(N)^3$-invariant general sextic model, the fixed-point structure in the long-range regime has remained unexplored so far. 

\medskip

In this paper, we first revisit the short-range fixed point analysis of this $O(N)^3$-invariant sextic model, using, as already mentioned above, a different method that the one used in \cite{jepsen2023rg, Fraser-Taliente:2024rql}.
Our method relies on the 
identification of the appropriate 
Feynman graphs, and on the computation of their Feynman amplitudes. This finally allows to obtain the $\beta$ functions of the model.
We thus obtain through this method the short-range fixed points already found in
\cite{jepsen2023rg, Fraser-Taliente:2024rql}.

Moreover, another result of our paper 
is the analysis of the fixed point structure of this model in the long-range case.  
We find that no additional fixed points appear with respect to those already found in the $U(N)^3-$invariant model.

\medskip

The paper is organized as follows. Section~\ref{sec2} recalls the 
$O(N)^3$-invariant general sextic model,
and recalls its dominant graphs in the large $N$ limit. In Section~\ref{sec3}, we perturbatively compute the $2$-point function and we determine the wavefunction renormalization. Section~\ref{sec4} is the central section of the paper and it derives the $\beta$ functions and the fixed points, in the short-range and the long-range cases.
An analysis of their stability is also performed. 
We compare our results with the ones of \cite{jepsen2023rg, Fraser-Taliente:2024rql} in the short-range case. In the long-range case, we compare our results with the ones
obtained for the $U(N)^3$-invariant model.
We give some final comments and perspective in Section \ref{sec5}.

\section{The model}
\label{sec2}
\subsection{The action of the model; short and long-range propagators}
\label{Section sextic}
\begin{figure}[H]
    \centering
    \includegraphics[width=0.9\textwidth]{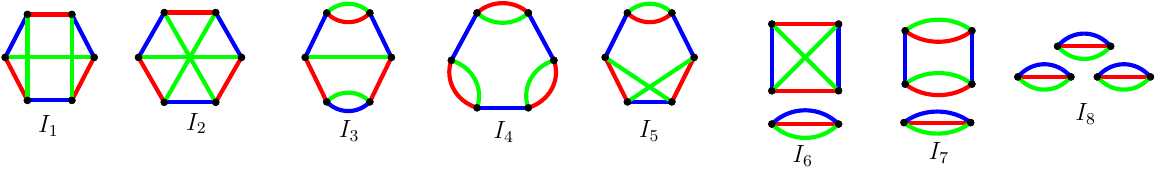}
    \caption{The $O(N)^3$-invariant sextic interactions. Color permutations are implicit.}
    \label{fig:sextic_bubbles}
\end{figure}

The partition function for the $O(N)^3$-invariant sextic tensor field theory in $d$ dimensions is
\begin{equation}
    \cZ=\int[dT]e^{-S[T]},
\end{equation}
where the $d$-dimensional action is given by
\begin{equation}
    S= \int \text{d}^d x \, T_{abc}(-\Delta)^\zeta T_{abc} + \sum_{b=1}^8 \frac{\lambda_b N^{-\alpha_b}}{6} I_b(T), 
    \label{action}
\end{equation}
where $T_{abc}$ denotes a real rank-3 tensor field with indices running from $1$ to $N$, $\lambda_b$ are the bare coupling constants, $I_b(T)$ are the $O(N)^3$-invariant interaction terms (represented graphically as bubble graphs, see Fig.~\ref{fig:sextic_bubbles}), and the coefficien $\alpha_b$ is the scaling of the respective interaction.

The sextic interactions $I_b(T)$ are constructed to be invariant under the $O(N)^3$ symmetry group, which acts on the tensor field as
\begin{equation}
    T_{a_1a_2a_3}\rightarrow R^{(1)}_{a_1b_1}R^{(2)}_{a_2b_2}R^{(3)}_{a_3b_3}T_{b_1b_2b_3},
\end{equation}
where $R^{(i)} \in O(N)$ are orthogonal matrices in the fundamental representation.

The eight independent interaction terms are listed in Fig.~\ref{fig:sextic_bubbles}. 
These interaction terms have the general form
\begin{equation}
    I_b(T)=\delta^{(b)}_{\ba\bb\bc\bd\be\bf}T_\ba T_\bb T_\bc T_\bd T_\be T_\bf, 
\end{equation}
where $\ba=(a_1a_2a_3)$. The eight indices contractions, symmetrized under color permutations, but not under external indices permutations, are
\begin{equation}
    \begin{split}
        \delta^{(1)}_{\ba\bb\bc\bd\be\bf}&=\delta_{a_1b_1}\delta_{b_2c_2}\delta_{c_1d_1}\delta_{d_2e_2}\delta_{e_1f_1}\delta_{f_2a_2}\delta_{a_3e_3}\delta_{b_3d_3}\delta_{f_3c_3},\\
        \delta^{(2)}_{\ba\bb\bc\bd\be\bf}&=\delta_{a_1b_1}\delta_{b_2c_2}\delta_{c_1d_1}\delta_{d_2e_2}\delta_{e_1f_1}\delta_{a_2f_2}\delta_{a_3d_3}\delta_{b_3e_3}\delta_{c_3f_3},\\
        \delta^{(3)}_{\ba\bb\bc\bd\be\bf}&=\frac{1}{3}(\delta_{a_1f_1}\delta_{a_2b_2}\delta_{a_3b_3}\delta_{b_1c_1}\delta_{c_2f_2}\delta_{c_3d_3}\delta_{e_3f_3}\delta_{d_1e_1}\delta_{d_2e_2}+(1\leftrightarrow 2)+(1\leftrightarrow 3)),\\
        \delta^{(4)}_{\ba\bb\bc\bd\be\bf}&=\frac{1}{3}\sum_{i=1}^3\delta_{a_ib_i}\delta_{c_id_i}\delta_{e_if_i}\prod_{j\neq i}\delta_{b_jc_j}\delta_{d_je_j}\delta_{a_jf_j},\\
        \delta^{(5)}_{\ba\bb\bc\bd\be\bf}&=\frac{1}{3}(\delta_{a_1f_1}\delta_{a_2b_2}\delta_{a_3b_3}\delta_{b_1c_1}\delta_{c_2e_2}\delta_{c_3d_3}\delta_{d_1e_1}\delta_{e_3f_3}\delta_{d_2e_2}+(1\leftrightarrow2)+(1\leftrightarrow3)),\\
        \delta^{(6)}_{\ba\bb\bc\bd\be\bf}&=\delta_{a_1b_1}\delta_{a_2c_2}\delta_{a_3d_3}\delta_{b_2d_2}\delta_{b_3c_3}\delta_{c_1d_1}\delta_{\be\bf},\\
          \delta^{(7)}_{\ba\bb\bc\bd\be\bf}&=\frac{1}{3}(\delta_{a_1b_1}\delta_{a_2b_2}\delta_{b_3c_3}\delta_{c_1d_1}\delta_{c_2d_2}\delta_{a_3d_3}+(1\leftrightarrow2)+(1\leftrightarrow3))\delta_{\bf\be},\\   
        \delta^{(8)}_{\ba\bb\bc\bd\be\bf}&=\delta_{\ba\bb}\delta_{\bc\bd}\delta_{\be\bf},
    \end{split}
\end{equation}
where $\delta_{\ba\bb}=\prod_{i=1}^3\delta_{a_ib_i}$

Among these eight invariant interaction terms, $I_1$ is commonly called the prismatic interaction, while $I_2$ is known as the wheel interaction. 

Let us also note that
the set of $O(N)^3$-invariant interactions is strictly larger than the subset of $U(N)^3$-invariant interactions: the latter must satisfy a bipartiteness constraint on the corresponding bubble graphs, which is absent for $O(N)^3$ symmetry. Consequently, the interactions $I_1$, $I_5$, and $I_6$ are allowed under $O(N)^3$-invariance but are forbidden under $U(N)^3$-invariance.

Note that
the kinetic term in eq.~\eqref{action} features a fractional power $\zeta$,  with $\zeta\leq1$. 
For $\zeta=1$, this reduces to the standard local (short-range) propagator. In this case, the sextic interactions become marginal at the upper critical dimension $d=3$. 
In the sequel, we analyze the short-range model using dimensional regularization in $d=3-\varepsilon$.

Long-range propagation, which corresponds to $0<\zeta<1$, thus involves a fractional Laplacian~\cite{Kwa_nicki_2017}. Such non-local operators are naturally defined in Fourier space, where the free propagator takes the form
\begin{equation}
    G_0(p)=\frac{1}{p^{2\zeta}}.
\end{equation}
From an RG perspective, tuning $\zeta$ allows the interactions to be rendered marginal in arbitrary dimensions, potentially yielding RG fixed points in any dimension~\cite{Trinchero_2019}. We study the long-range model in $d\leq3$ with the parametrization $\zeta=\frac{d+\varepsilon}{3}$, which ensures marginality in the limit $\varepsilon\rightarrow0$.

Recall that for the $U(N)^3$-invariant interactions \cite{Benedetti_2020}, Wilson-Fisher type fixed points are found in the short-range, while a set of fixed-point parameterized by the wheel coupling is found in the long-range.
In both short and long-range, the sextic couplings have dimension $[\lambda_i]=2\varepsilon$.\\

\begin{figure}[H]
    \centering
    \includegraphics[scale=0.7]{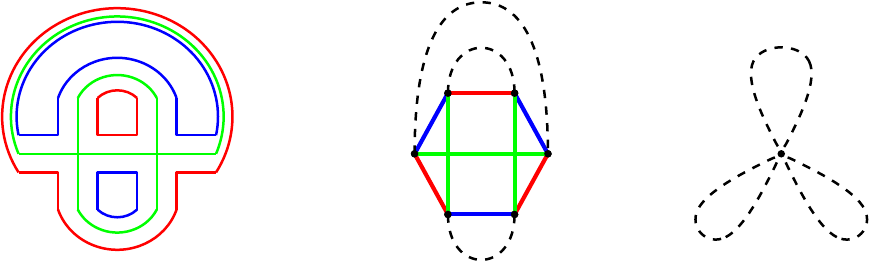}
    \caption{A triple-tadpole graph displayed in the stranded (left), bubble (center), and Feynman (right) representations.}
    \label{fig:rpz}
\end{figure}

Feynman graphs arising from the perturbative expansion of tensor models can be represented in multiple equivalent ways, each highlighting different aspects of the tensorial structure. We illustrate these representations using a triple-tadpole graph (a graph with three tadpole loops attached to the same sextic interaction vertex) in Fig.~\ref{fig:rpz}. 
\begin{itemize}
    \item The \textit{stranded representation} assigns a distinct color to each tensor index and explicitly depicts the propagation of colored strands through the graph (Fig.~\ref{fig:rpz}, left).
    \item The \textit{bubble representation} associates interaction vertices with $3$-colored graphs (bubbles) encoding the internal index structure, while propagators are represented by dashed edges connecting these bubbles (Fig.~\ref{fig:rpz}, center).
    \item The \textit{Feynman representation} suppresses all explicit index structure, representing interaction vertices as dots and propagators as dashed edges (Fig.~\ref{fig:rpz}, right). 
\end{itemize}
A \textit{ribbon jacket} $\mathcal{J}_i(\mathcal{G})$, with $i\in\{1,2,3\}$, is the graph obtained by removing all strands of color $i$ from the stranded representation of a graph $\mathcal{G}$. Each ribbon jacket is itself a ribbon graph.

The model~\eqref{action} was previously studied in \cite{Giombi_2018} using scalings that favor the prismatic interaction, a choice motivated by the analysis of the renormalization group flow in the prismatic sector. As already mentioned above, in the present paper we use instead the optimal scalings introduced in \cite{Carrozza_2016}. These scalings are given by
\begin{equation}
    \alpha_b=3+\tfrac{1}{2} \sum_l \delta_l^{(b)},
\end{equation}
where $\delta_l^{(b)} = |J_l^{(b)}| - 1$, and $|J_l^{(b)}|$ denotes the number of connected components in the $l$-th jacket of bubble $b$. 

The optimal scalings take the explicit values
\begin{equation}
    \alpha_1=\alpha_2=3, \quad \alpha_{3}=\alpha_4 = 4, \quad \alpha_5 =7/2, \quad \alpha_6 = 9/2, \quad \alpha_7 = 5, \quad \alpha_8 = 6.
\end{equation}
This choice of scalings is consistent with the one adopted in \cite{jepsen2023rg,Fraser-Taliente:2024rql} for the study of $O(N)^3$-invariant interactions.

\subsection{Dominant graphs of the model}

\label{SectionLargeN}
The $1/N$ expansion of tensor models is controlled by the degree $\omega$, see for example \cite{gurau_complete_2012}. The degree of a Feynman graph $\mathcal{G}$ is defined as
\begin{equation}
    \omega(\mathcal{G})=3+\sum_{b\in V(\mathcal{G})}\alpha_b-F(\mathcal{G}),
\end{equation}
where the sum runs over all interaction vertices in $\mathcal{G}$ and $F(\mathcal{G})$ denotes the number of faces of the graph. Recall that a face of a tensor graph is defined as a colored cycle in the \textit{stranded representation}. Equivalently, in the \textit{bubble representation}, a face corresponds to a closed cycle formed by the alternating sequence of dashed edges and colored edges.

Vacuum graphs that dominate in the large $N$ limit have vanishing degree, $\omega=0$. For the $U(N)^3$ model, the dominant vacuum graphs are the so-called  \textit{melon-tadpole} graphs \cite{bonzom2015colored, Benedetti_2020}. In the case of the $O(N)^3$ sextic model, the class of dominant graphs in the large $N$ limit 
has been identified in the companion paper 
\cite{largeN}. 

As already mentioned above, in this paper we study perturbative RG flows in the massless $O(N)^3$ sextic model. 
Since we are using the dimensional regularization scheme, 
Feynman integrals of tadpole graphs identically vanish and will therefore be omitted from our analysis.

The dominant $2$-point graphs at large $N$ have degree $\omega=3$. 
This is a direct consequence of the fact that, in order to obtain a $2-$point dominant graph from a dominant vacuum graph, one needs to 'cut' three faces of the vacuum graph, thus leading to an increase by $3$ of the degree.

\begin{figure}[H]
    \centering
    \includegraphics[width=0.8\linewidth]{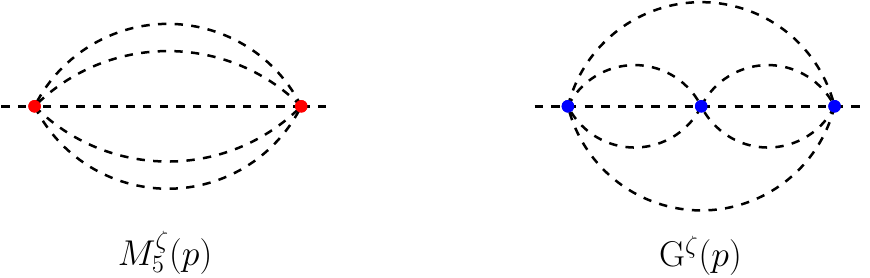}
    \caption{Dominant $2-$point graphs up to cubic order in $\lambda$.}
    \label{fig:2p_dom}
\end{figure}

Recall
from \cite{largeN} that,
at order $\mathcal{O}(\lambda^3)$, the dominant $2$-point contributions are given by the \textit{melon} graph $M_5^\zeta$ and the \textit{glasses} graph $\mathrm{G}^\zeta$, depicted in Fig.~\ref{fig:2p_dom}. In the figure, propagators are represented as dashed lines,
red vertices are either both $\lambda_1$ or both $\lambda_2$ vertices, and blue vertices are $\lambda_1$ vertices.

Let us discuss here the melonic character of the Feynman graphs $M_5^\zeta$ and $\mathrm{G}^\zeta$ of Fig. \ref{fig:2p_dom}. If one places himself in a single field model setting, than the graph $M_5^\zeta$ is a melonic one (with a standard iterative structure), while the graph $\mathrm{G}^\zeta$ is a non-melonic one. However, in a two-field theory, this graph also has the standard iterative melon stucture.

The $6$-point graphs generate the sextic interactions $I_b(T)$ through radiative corrections. The color indices propagation in a $6-$point graph matches the structure of one of the sextic interactions. A $6$-point graph $\mathcal{G}_b$ that generates $I_b(T)$ is dominant in the large $N$ limit if its degree satisfies $\omega(\mathcal{G}_b)=3+\alpha_b$. Such graph $\cG_b$ scales exactly as the interaction it generates, and is therefore dominant in the large $N$ limit. 

\begin{figure}[H]
    \centering
    \includegraphics[scale=0.7]{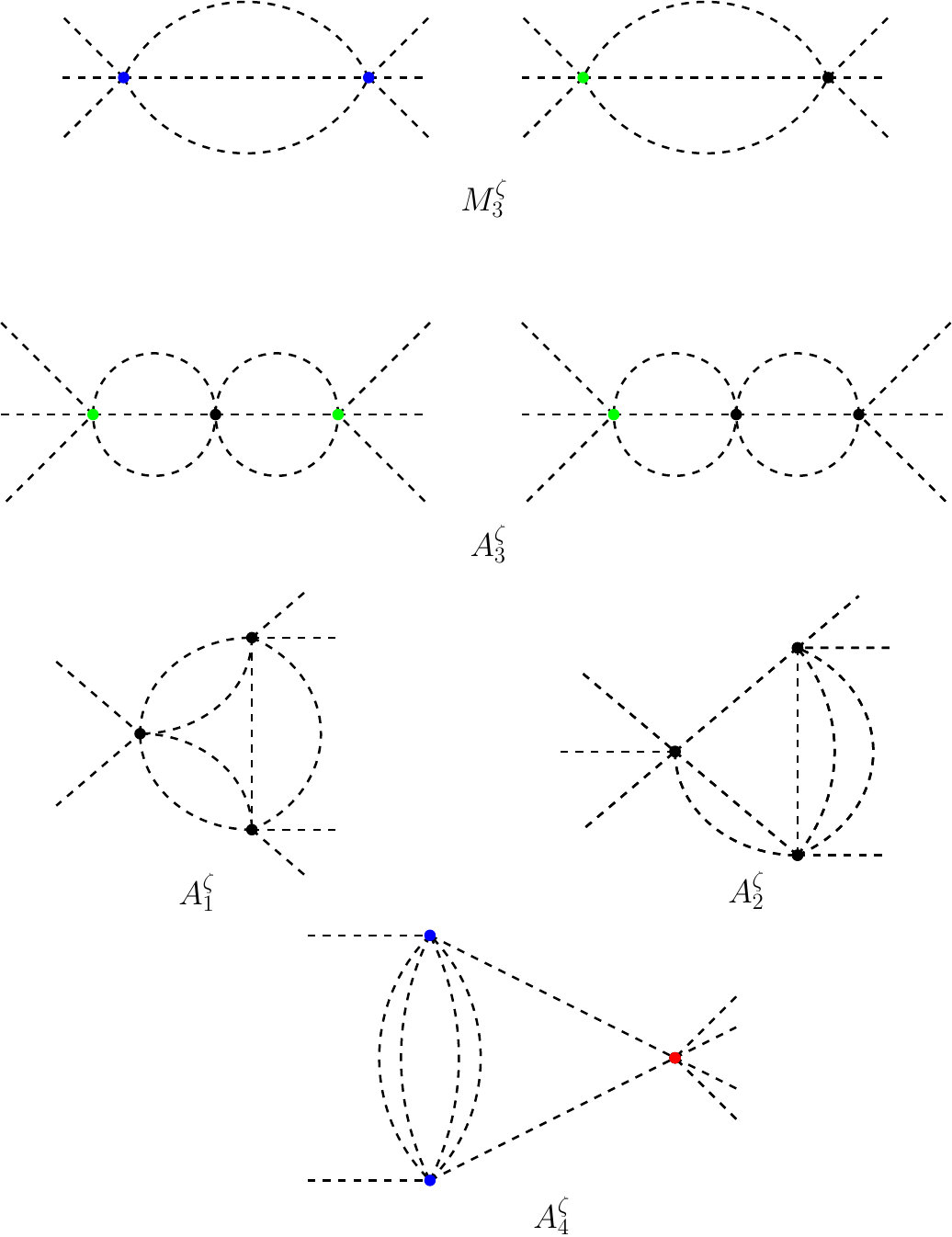}
    \caption{Leading order $6-$point graphs up to cubic order in $\lambda$.}
    \label{fig:6p_dom}
\end{figure}

The dominant $6$-point graphs at order $\mathcal{O}(\lambda^3)$ are shown in Fig.~\ref{fig:6p_dom}. In the figure,
blue vertices can be both $\lambda_1$ or both $\lambda_2$ vertices; black vertices are $\lambda_1$ vertices; green vertices can be $\lambda_5$ or $\lambda_6$ and, finally, the red vertex can be any of the eight types of vertices.

Let us also mention  that all these dominant graphs appear in the next-to-leading order expansion of the $U(N)^3$ sextic model as well, see \cite{Harribey_2022}.

Note that the dominant graphs in Fig.~\ref{fig:2p_dom} and Fig.~\ref{fig:6p_dom} are represented in the \textit{Feynman representation}, where the internal tensorial structure at vertices is omitted. This representation is well-suited for computing Feynman loop integrals associated to these Feynman graphs. To analyze the internal tensorial structure explicitly, the \textit{bubble representation} is more appropriate. 

\begin{figure}[H]
    \centering
    \includegraphics[scale=0.7]{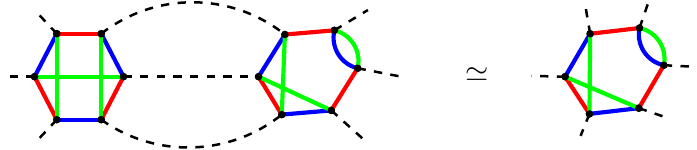}
    \caption{Dominant $6-$point melon graph with $I_5$ external index structure. }
    \label{fig:I5fromM3}
\end{figure}

\begin{figure}[H]
    \centering
    \includegraphics[scale=0.7]{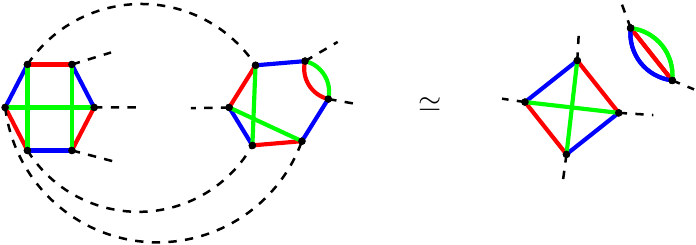}
    \caption{Dominant $6-$point melon graph with $I_6$ external index structure.}
    \label{fig:I6fromM3}
\end{figure}

As an illustration, we present in Fig.~\ref{fig:I5fromM3} and Fig.~\ref{fig:I6fromM3} two melon graphs $M^\zeta_3$ built on the interactions $I_1$ and $I_5$, which are represented using the bubble representation, in order to make the index contraction structure manifest. 


The melon graph in Fig.~\ref{fig:I5fromM3} generates the interaction $I_5$ at two loops. Its degree is 
\begin{equation}
    \omega=3+\alpha_1+\alpha_5-3=3+\alpha_5.
\end{equation}
This graph is therefore dominant in the large $N$ limit. 

The graph in Fig.~\ref{fig:I6fromM3} generates instead the interaction $I_6$ at two loops. This is due to a different pattern of index contractions. The  degree of this graph is:
\begin{equation}
    \omega=3+\alpha_1+\alpha_5-2=3+\alpha_6. 
\end{equation}
This graph is thus also dominant in the large $N$ limit. This leads to the fact that the $6-$point vertex functions $\Gamma_5^{(6)}$ and $\Gamma_6^{(6)}$ both contain $\lambda_5\lambda_6M_3^\zeta$.

\section{Wavefunction renormalization}
\label{sec3}

\begin{figure}[H]
    \centering
    \includegraphics[width=1\linewidth]{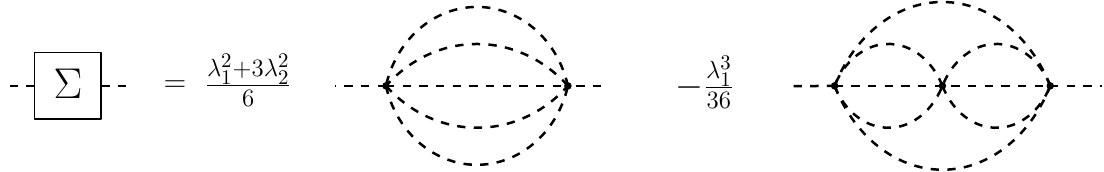}
    \caption{Perturbative expansion of the self-energy $\Sigma$ up to order $\cO(\lambda^3)$}
    \label{fig:pert_sigma}
\end{figure}

In this section, we compute the $2-$point function RG flow. 
The $2-$point function is given by the Schwinger-Dyson Equation (SDE)
\begin{equation}
    \Gamma^{(2)}(p)=p^{2\zeta}-\Sigma(p), 
\end{equation}
where $\Sigma(p)$ is the self-energy which is given by the sum over the $2-$point 1PI graphs. As already mentioned above, 
the first dominant perturbative contributions to $\Sigma$ are the graphs $M_5^\zeta$ and $G^\zeta$. 
The diagrammatic SDE at order $\lambda^3$ is given in Fig.~\ref{fig:pert_sigma}.

In momentum space this writes
\begin{equation}
   \Gamma^{(2)}(p)=p^{2\zeta}-\frac{1}{6}(\lambda_1^2+3\lambda_2^2)M_5^\zeta(p)+\frac{1}{36}\lambda_1^3\text{G}^\zeta(p).
\end{equation}

The wavefunction renormalization $Z$ is defined by
\begin{equation}
    \Gamma^{(2)}_R(p)=Z\, \Gamma^{(2)}(p),
\end{equation}
where $ \Gamma^{(2)}_R(p)$ is the renormalized $2-$point function and  $\Gamma^{(2)}(p)$ the bare 
$2-$point function. 

The renormalization conditions for a massless theory with a fractional Laplacian are
\begin{align}
    &\Gamma^{(2)}_R(p=0)=0,\\
    &\frac{\del\Gamma^{(2)}_R}{\del p^{2\zeta}}|_{p=\mu}=1,
    \label{renormalization condition}
\end{align}
where $\mu$ is a typical energy scale. 
Applying the renormalization condition and solving iteratively in $\lambda$ gives
\begin{equation}
    Z=1+\frac{(\lambda_1^2+3\lambda_2^2)}{6}\frac{\partial M_5^\zeta(p)}{\partial p^{2\zeta}}\bigg|_{p=\mu}-\frac{\lambda_1^3}{36} \frac{\partial \text{G}^\zeta(p)}{\partial p^{2\zeta}}\bigg|_{p=\mu}.
    \label{Zbare}
\end{equation}
Note that in eq.~\eqref{Zbare} above the wavefunction renormalization $Z$ is expressed as a function of the bare couplings $\lambda_1,~\lambda_2$. 

Let us now denote by $\{g_i\}$ the renormalized coupling constants.
The bare couplings $\{\lambda_i\}$ are then related to the renormalized ones $\{g_i\}$ by the following relations 
\begin{equation}
    g_{i}=\mu^{-2\varepsilon}Z^3\Gamma^{(6)}_i(\{p_j\},\mu,\{\lambda\}),
    \label{renormalized_couplings}
\end{equation}
where $\Gamma^{(6)}_i$ is the $6-$point function for the interaction term $I_i$, and the set 
$\{p_j\}$ is the set of external momenta.

\begin{figure}[H]
    \centering
    \includegraphics[width=\textwidth]{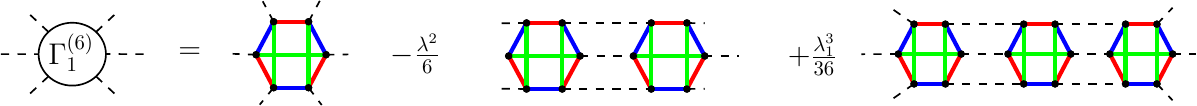}
    \caption{Perturbative expansion of the prismatic $6-$point function $\Gamma_1^{(6)}$. }
    \label{fig:gamma1}
\end{figure}

\begin{figure}[H]
    \centering
    \includegraphics{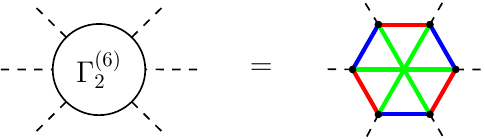}
    \caption{Perturbative expansion of the wheel $6-$point function $\Gamma_1^{(6)}$}
    \label{fig:gamma2}
\end{figure}

The $6-$point function for the prismatic interaction and the wheel interaction are obtained, as usually, by diagrammatic perturbative expansion. We exhibit this diagrammatic expansion of $\Gamma_1^{(6)}$ and $\Gamma_2^{(6)}$ on Fig. \ref{fig:gamma1} and Fig. \ref{fig:gamma2} (see \cite{Krajewski_2023} for the recursive relation for $\Gamma_1^{(6)}$). Note that, in the case of the wheel interaction, see Fig.  \ref{fig:gamma2}, there are no radiative correction up to order $3$ perturbation theory.

The $6-$point  functions above then writes:
\begin{equation}
\begin{split}
    &\Gamma_1^{(6)}=\lambda_1-\frac{\lambda_1^2}{6}M_3^\zeta(\mu)+\frac{\lambda_1^3}{36}A_3(\mu),\\
    &\Gamma_2^{(6)}=\lambda_2.
    \label{vertex1and2}
\end{split}
\end{equation}

Eq.~\eqref{renormalized_couplings} can therefore be perturbatively inverted to obtain the expansion of the bare coupling constants $\lambda_1$ and $\lambda_2$ 
as a function of 
the renormalized coupling constants $g_1$ and $g_2$:
\begin{align}
    \lambda_2&=\mu^{2\varepsilon}g_2+\cO(g^4),  \label{invg1}\\
    \lambda_1&=\mu^{2\varepsilon}g_1+\frac{\mu^{4\varepsilon}g_1^2}{6}M_3^\zeta(\mu)+\cO(g^3).
    \label{invg1g2}
\end{align}
The wavefunction renormalization then writes as a function of the renormalized couplings $g_1$ and $g_2$ as:
\begin{equation}
    Z=1+\frac{(g_1^2+3g_2^2)}{6}\mu^{4\varepsilon}\frac{\partial M_5^\zeta(p)}{\partial p^{2\zeta}}|_{p=\mu}+\frac{2\mu^{6\varepsilon}g_1^3}{36}\frac{\partial M_5^\zeta(p)}{\partial p^{2\zeta}}|_{p=\mu}M_3^\zeta(\mu)-\frac{\mu^{6\varepsilon}g_1^3}{36}\frac{\partial \text{G}^\zeta(p)}{\partial p^{2\zeta}}|_{p=\mu}.
    \label{Zgeneric}
\end{equation}
For the sake of completeness, the general melonic integral is explicitly computed in Appendix \ref{Appint} 
(see also \cite{Fraser-Taliente:2024rql}). 
Using eq.~\eqref{generalmelon}, 
the Feynman integrals of the melonic graphs $M_3^\zeta$ and resp. $M_5^\zeta$
are
\begin{equation}
\label{M3}
    M^\zeta_3(p)=\frac{p^{2d-6\zeta}}{(4\pi)^d}\frac{\Gamma(\frac{d}{2}-\zeta)^3}{\Gamma(\zeta)^3}\frac{\Gamma(3\zeta-d)}{\Gamma(\frac{3d}{2}-3\zeta)},
\end{equation}
and resp. 
\begin{equation}
    M^\zeta_5(p)=\frac{p^{4d-10\zeta}}{(4\pi)^{2d}}\frac{\Gamma(\frac{d}{2}-\zeta)^5}{\Gamma(\zeta)^5}\frac{\Gamma(5\zeta-2d)}{\Gamma(\frac{5d}{2}-5\zeta)}.
    \label{M5}
\end{equation}
The Feynman integral  $\text{G}^\zeta(p)$ is also explicitly computed in Appendix~\ref{Appint}. One gets:
\begin{equation}
\label{glasses}
\text{G}^\zeta(p)=\frac{1}{(4\pi)^{3d}}\frac{\Gamma(\frac{d}{2}-\zeta)^8}{\Gamma(\zeta)^8}\frac{\Gamma(3\zeta-d)^2}{\Gamma(\frac{3d}{2}-3\zeta)^2}\frac{\Gamma(\frac{5d}{2}-6\zeta)}{\Gamma(6\zeta-2d)}\frac{\Gamma(8\zeta-3d)}{\Gamma(\frac{7d}{2}-8\zeta)}\frac{1}{p^{2(8\zeta-3d)}}.
\end{equation}

\subsection{Short-range}

In order to investigate the short-range behavior 
of the $2-$point function, we set 
$d=3-\varepsilon$ and $\zeta=1$ in equations \eqref{M3}, \eqref{M5} and resp. \eqref{glasses}. One gets: 
\begin{equation}
    M^{\zeta=1}_3(p)=\frac{p^{-2\varepsilon}}{(4\pi)^3}\frac{\pi}{2}\left(\frac{1}{\varepsilon}-\gamma+\cO(\varepsilon^1)\right),
\end{equation}
\begin{equation}
M^{\zeta=1}_5(p)=p^{2(1-2\varepsilon)}\frac{4\pi^2}{3(4\pi)^6}\left(-\frac{1}{2\varepsilon}+(\gamma-1)+\mathcal{O}(\varepsilon^1)\right), 
\end{equation}
and resp.
\begin{equation}
\text{G}^{\zeta=1}(p)=\frac{p^{2(1-3\varepsilon)}}{(4\pi)^9}\frac{8\pi^3}{3}\left(-\frac{2}{3\varepsilon^2}+\frac{2(\gamma-1)}{\varepsilon}+\mathcal{O}(\varepsilon^0)\right),
    \label{nonlocal}
\end{equation}
where $\gamma$ is the Euler-Mascheroni constant.

Inserting these results in 
eq.~\eqref{Zgeneric} gives the short-range wavefunction renormalization:
\begin{equation}
\label{Zshort}
Z=1-(g_1^2+3g_2^2)\frac{2\pi^2}{3(4\pi)^2}\frac{1}{\varepsilon}+\frac{8g_1^3}{3}\frac{2\pi^2}{3(4\pi)^9}\frac{1}{\varepsilon}.
\end{equation}
Note that, if one sets $g_1=0$, one recovers the short-range wave function renormalization of the $U(N)^3-$invariant model:
\begin{equation}
Z=1-3g_2^2\frac{2\pi^2}{3(4\pi)^2}\frac{1}{\varepsilon}.
\end{equation}

Finally, let us note that the wavefunction renormalization 
\eqref{Zshort}
diverges in the short-range model in the limit  $\varepsilon\rightarrow0$.   
The anomalous dimension $\eta= \mu\partial_\mu \log Z$ is 
\begin{equation}
    \eta=\frac{8\pi^2}{3}(g_1^2+3g_2^2)+\cO(g^3). 
\end{equation}

\subsection{Long-range}
In order to investigate the long-range behavior 
of the $2-$point function, we set 
$\zeta=\frac{d+\varepsilon}{3}$, with $d<3$. One gets:
\begin{equation}
    M_3^{\zeta=\frac{d+\varepsilon}{3}}(p)=\frac{p^{-2\varepsilon}}{(4\pi)^d}\frac{\Gamma(\frac{d}{6})^3}{\Gamma(\frac{d}{3})^3\Gamma(\frac{d}{2})}\left(\frac{1}{\varepsilon}-\gamma+\cO(\varepsilon)\right),
\end{equation}
\begin{equation}
M_5^{\zeta=\frac{d+\varepsilon}{3}}(p)=\frac{p^{2\zeta(1-2\varepsilon/\zeta)}}{(4\pi)^{2d}}\frac{\Gamma(\frac{d}{6})^5}{\Gamma(\frac{d}{3})^5}\frac{\Gamma(-\frac{d}{3})}{\Gamma(\frac{5d}{6})},
\end{equation}
and
\begin{equation}
\text{G}^{\zeta=\frac{d+\varepsilon}{3}}(p)=\frac{p^{2\zeta(1-3\varepsilon/\zeta)}}{(4\pi)^{3d}}\frac{\Gamma(\frac{d}{6})^8}{\Gamma(\frac{d}{3})^8}\frac{\Gamma(-\frac{d}{3})}{\Gamma(\frac{d}{2})\Gamma(\frac{5d}{6})}\frac{2}{\varepsilon}    +\mathcal{O}(\varepsilon^0).
\end{equation}
Let us denote by $\cZ$ the long-range wavefunction renormalization, which now becomes:
\begin{equation}
\label{Zlong}
    \cZ=1-\frac{g_1^2+3g_2^2}{6(4\pi)^{2d}}\frac{\Gamma(\frac{d}{6})^5\Gamma(-\frac{d}{3})}{\Gamma(\frac{d}{3})^5\Gamma(\frac{5d}{6})}+\frac{g_1^3}{d(4\pi)^{3d}}\frac{\Gamma(\frac{d}{6})^8\Gamma(-\frac{d}{3})}{\Gamma(\frac{d}{3})^8\Gamma(\frac{d}{2})\Gamma(\frac{5d}{6})}. 
\end{equation}
Note that if one sets $g_1=0$, the expression above simplifies to:
\begin{equation}
    \cZ=1-\frac{g_2^2}{2(4\pi)^{2d}}\frac{\Gamma(\frac{d}{6})^5\Gamma(-\frac{d}{3})}{\Gamma(\frac{d}{3})^5\Gamma(\frac{5d}{6})}. 
\end{equation}
We thus recover, as expected, the $U(N)^3$ model wave function renormalization, see again  \cite{Benedetti_2020}.

Let us end this section, by noticing that the long-range wave function renormalization
\eqref{Zlong}
is finite for $d<3$.
The wavefunction renormalization is therefore a finite, $\varepsilon$-independent constant in the long-range model. 
Consequently, the anomalous dimension $\eta = \mu \partial_\mu \log \cZ$ vanishes, as 
expected for long-range models (see, for example \cite{Benedetti_2019}).

\section{$\beta$-functions and fixed points}
\label{sec4}

In this section we compute the $\beta$-functions, we identify the fixed points of the our model and we compare our results with the ones of the $U(N)^3$ invariant model \cite{Benedetti_2020}. 

Recall that the renormalized couplings $g_i$ write in term of the bare couplings $\lambda_i$ as
\begin{equation}
    g_{i}=\mu^{-2\varepsilon}Z^3\Gamma^{(6)}_i(\{p_j\},\mu,\{\lambda_i\}),
    \label{renormalized couplings}
\end{equation}
where $\Gamma^{(6)}_i$ denote, as already mentioned above, the $1$PI $6-$point function for the $i$-th interaction, evaluated at external momenta $\{p_j\}$. For the $6-$point functions computations, we use BPHZ subtraction at zero momentum, together with the modified propagator for IR divergences regulation (see \cite{Benedetti:2020rrq, Harribey_2022})
\begin{equation}
    G_0^{(\mu)}=\frac{1}{(p^2+\mu^2)^\zeta},
\end{equation}
where $\mu$ is a typical energy scale. 
The $6-$point functions $\Gamma^{(6)}_i$ are obtained again through a diagrammatic expansion in the bubble representation, which makes the external index structure explicit. In what follows, we suppress the superscript $\zeta$ on the melon integral $M_3^\zeta$ and the auxiliary integrals $A_i^\zeta$ to simplify notation.

The $6-$point functions write,
as expressions of the sextic Feynman integrals of the graphs of Fig.~\ref{fig:6p_dom}, as:
\begin{equation*}
\begin{split}
    &\Gamma_1^{(6)}=\lambda_1-\frac{\lambda_1^2}{6}M_3(\mu)+\frac{\lambda_1^3}{36}A_3(\mu),\\
    &\Gamma_2^{(6)}=\lambda_2,\\
    &\Gamma_3^{(6)}=\lambda_3-\left(2\lambda_1^2+9\lambda_2^2+\frac{2\lambda_5^2}{9}\right)M_3(\mu)+\frac{2\lambda_1^3}{3}A_2(\mu)\\
    &+\left(2\lambda_1^3+6\lambda_1^2\lambda_2+6\lambda_1\lambda_2^2+18 \lambda_2^3+\frac{2\lambda_1^2\lambda_3}{3}+2\lambda_2^2\lambda_3\right)A_4(\mu)+\frac{2\lambda_1\lambda_5^2}{27}A_3(\mu),\\
    &\Gamma_4^{(6)}=\lambda_4-\frac{\lambda_5^2}{9}M_3(\mu)+\frac{\lambda_1^3}{9}A_1(\mu)+(\lambda_1^2+3\lambda_2^2)\lambda_4A_4(\mu)+\lambda_1\lambda_5^2\frac{(2A_2(\mu)+A_3(\mu))}{27},\\
    &\Gamma_5^{(6)}=\lambda_5-\frac{2\lambda_1\lambda_5}{3} M_3(\mu)+\frac{\lambda_1^2\lambda_5}{3}A_3(\mu)+\left(\frac{\lambda_1^2\lambda_5}{9}+\lambda_2^2\lambda_5\right)A_4(\mu),\\
    &\Gamma_6^{(6)}=\lambda_6-\frac{4\lambda_1\lambda_6}{3} M_3(\mu)+\frac{3\lambda_1^2\lambda_6}{4}A_3(\mu)-\frac{2\lambda_1\lambda_5}{3}M_3(\mu)+\frac{\lambda^2_1\lambda_5}{2}A_3(\mu)+\\
    &\left(\frac{4\lambda_1^2\lambda_5}{3}+\frac{5\lambda_1^2\lambda_6}{3}+4\lambda_2^2\lambda_5+5\lambda_2^2\lambda_6\right)A_4(\mu),\\
    &\Gamma_7^{(6)}=\lambda_7-\left(\lambda_1^2+\frac{7\lambda_5^2}{9}+\frac{4\lambda_5\lambda_6}{3}\right)M_3(\mu)+\frac{2\lambda_1^3}{3}A_1(\mu)+\frac{\lambda_1^3}{3}A_2(\mu)+\\ &\left(3\lambda_1^3+9\lambda_1^2\lambda_2+\frac{10\lambda_1^2\lambda_3}{3}+4\lambda_1^2\lambda_4+9\lambda_2^2\lambda_1+27\lambda_2^3+10\lambda_2^2\lambda_3+12\lambda_2^2\lambda_4+\frac{7\lambda_1^2\lambda_7}{3}+7\lambda_2^2\lambda_7\right)A_4(\mu)\\
    &+\frac{16\lambda_1\lambda_5\lambda_6}{9}(A_2(\mu)+2A_3(\mu))+\frac{52\lambda_1\lambda_5^2}{27}A_2(\mu),\\
    & \Gamma_8^{(6)}=\lambda_8-\left(\lambda_2^2+\frac{2\lambda_6^2}{9}+\frac{4\lambda_5\lambda_6}{3}+\frac{2\lambda_5^2}{9}\right)M_3(\mu)+\frac{2\lambda_1^3}{9}A_1(\mu)+\left(\frac{220\lambda_1\lambda_5\lambda_6}{81}+\frac{8\lambda_1\lambda_6^2}{3}\right)A_2(\mu)\\
    &+\frac{10}{27}\lambda_1(\lambda_6^2+2\lambda_5^2)A_3(\mu)+\left(\lambda_1^2\lambda_3+\frac{8\lambda_1^2\lambda_7}{3}+5\lambda_1^2\lambda_8+3\lambda_2^2\lambda_3+8\lambda_2^2\lambda_7+15\lambda_2^2\lambda_8\right)A_4(\mu).
\end{split}
\end{equation*}

One can then prove that the $\beta$ functions, given by the general relation 
$$\beta_i=\mu\partial_\mu g_i,$$
further write
\begin{equation}
    \beta_i=(-2\varepsilon+3\eta)g_i+\mu^{-2\varepsilon}Z^3\mu\partial_\mu\Gamma^{(6)}_i(\mu,\{\lambda(g)\}), 
    \label{beta_generic}
\end{equation}
where, as already mentioned above, $\eta$ denotes the anomalous dimension of the field.

In order to express the $\beta$ functions in terms of the renormalized couplings $g_i$, one must invert the relations $\{\lambda(g)\}$ given in eq.~\eqref{renormalized couplings}. 
Recall that, for the couplings $g_1$ and $g_2$, the inverted relations are given in eq.~\eqref{invg1g2}. 

To implement this inversion for the remaining $6$ other coupling constants, we expand the bare couplings $\lambda_i$ in terms of the renormalized ones up to order $g^3$.
The general expression for this inversion writes
\begin{equation}
    \lambda_i=g_i\mu^{2\varepsilon}+a_{ijk}g_jg_k\mu^{4\varepsilon}+b_{ijkl}g_jg_kg_l \mu^{6\varepsilon}, 
    \label{lam(g)}
\end{equation}
where $a_{ijk}$ and $b_{ijkl}$ are general coefficients. 
 Inserting eq.~\eqref{lam(g)} into eq.~\eqref{renormalized couplings} and solving order by order in perturbation theory, leads to  the 
 explicit expressions of the 
 coefficients $a_{ijk}$ and $b_{ijkl}$.
Substituting then the explicit  expansion 
\eqref{lam(g)}
into eq.~\eqref{beta_generic} leads to
the $\beta$ functions expressed in terms of renormalized couplings $g_i$. 

These $\beta$ functions are explicitly written in Appendix~\ref{AppBeta} and the calculation is exemplified for the case of $\beta_4$ in Appendix~\ref{AppBeta4}. The computations for the other $\beta$ functions are analogous.

\subsection{Short-range}

In the short-range case  $\zeta=1$, the loop-integrals are (see Appendix $C$ of \cite{Harribey_2022} for the detailed computation using Mellin-Barnes representation):

\begin{equation}
    M_3^{\zeta=1}(\mu)=\mu^{-2\varepsilon}\frac{\pi}{(4\pi)^3}\left(\frac{2}{\varepsilon}+\psi\left(\frac{1}{2}\right)+\psi\left(\frac{3}{2}\right)+4\log\left(\frac{2}{3}\right) \right)+\cO(\varepsilon)
\end{equation}

\begin{equation}
    A_1^{\zeta=1}(\mu)=\mu^{-4\varepsilon}\frac{\pi^4}{(4\pi)^{6}}\left(\frac{1}{\varepsilon}+\cO(\varepsilon^0)\right),
\end{equation}

\begin{equation}
    A_2^{\zeta=1}(\mu)=\mu^{-4\varepsilon}\frac{2\pi^2}{(4\pi)^6}\left(\frac{1}{\varepsilon^2}+\frac{2}{\varepsilon}\left(2\log\left(\frac{2}{3}\right)+\psi\left(\frac{3}{2}\right)\right)+\cO(\varepsilon^0)\right),
\end{equation}

\begin{equation}
    A_3^{\zeta=1}(\mu)=\Big(M_3(\mu)^{\zeta=1}\Big)^2,
\end{equation}

\begin{equation}
    A_4^{\zeta=1}(\mu)=-\mu^{-4\varepsilon}\frac{2\pi^2}{\varepsilon(4\pi)^6}+\cO(\varepsilon^0).
\end{equation}
These integrals can therefore be inserted in the expressions \eqref{betageneric}
of the  $\beta$ functions.

We now use the following rescalings:
$$\Tilde{g}=g(4\pi)^3$$ and 
$$\tilde{\beta}=\frac{\beta}{(4\pi)^3}.$$
The short-range $\beta$-functions of our model then write (after forgetting the tilde):
\begin{equation}
    \begin{split}
        \beta_1&=\frac{8}{3} \, \pi^{2} g_{1}^{3} + 8 \, \pi^{2} g_{1} g_{2}^{2} + \frac{4}{3} \, \pi g_{1}^{2} - 2 \, \varepsilon g_{1},\\
        \beta_2&=8 \, \pi^{2} g_{2}^{3} + \frac{2}{3} \, {\left(4 \, \pi^{2} g_{1}^{2} - 3 \, \varepsilon\right)} g_{2},\\
        \beta_3&=-\frac{16}{3} \, \pi^{2} g_{1}^{3} + 48 \, \pi^{2} g_{1}^{2} g_{2} + 144 \, \pi^{2} g_{2}^{3} + 8 \, \pi g_{1}^{2} + 12 \, {\left(3 \, \pi + 4 \, \pi^{2} g_{1}\right)} g_{2}^{2}\\ &+ \frac{8}{27} \, {\left(3 \, \pi - 8 \, \pi^{2} g_{1}\right)} g_{5}^{2} + 2 \, {\left(4 \, \pi^{2} g_{1}^{2} + 12 \, \pi^{2} g_{2}^{2} - \varepsilon\right)} g_{3},\\    
        \beta_4&=-\frac{4}{9} \, \pi^{4} g_{1}^{3} + \frac{4}{27} \, {\left(3 \, \pi - 8 \, \pi^{2} g_{1}\right)} g_{5}^{2} + \frac{2}{3} \, {\left(16 \, \pi^{2} g_{1}^{2} + 48 \, \pi^{2} g_{2}^{2} - 3 \, \varepsilon\right)} g_{4},\\  
        \beta_5&=\frac{2}{9} \, {\left(8 \, \pi^{2} g_{1}^{2} + 72 \, \pi^{2} g_{2}^{2} + 12 \, \pi g_{1} - 9 \, \varepsilon\right)} g_{5},\\
        \beta_6&=-\frac{8}{9} \, {\left(8 \, \pi^{2} g_{1}^{2} - 36 \, \pi^{2} g_{2}^{2} - 3 \, \pi g_{1}\right)} g_{5} - \frac{2}{3} \, {\left(8 \, \pi^{2} g_{1}^{2} - 72 \, \pi^{2} g_{2}^{2} - 8 \, \pi g_{1} + 3 \, \varepsilon\right)} g_{6},\\
        \beta_7&= 72 \, \pi^{2} g_{1} g_{2}^{2} + 216 \, \pi^{2} g_{2}^{3} - \frac{8}{3} \, {\left(\pi^{4} - 5 \, \pi^{2}\right)} g_{1}^{3} + 4 \, \pi g_{1}^{2} + \frac{4}{27} \, {\left(21 \, \pi - 160 \, \pi^{2} g_{1}\right)} g_{5}^{2} \\
        &+ \frac{16}{9} \, {\left(3 \, \pi - 16 \, \pi^{2} g_{1}\right)} g_{5} g_{6} + \frac{80}{3} \,  {\left(\pi^{2} g_{1}^{2} + 3 \, \pi^{2} g_{2}^{2}\right)} g_{3} + 32 \, {\left(\pi^{2} g_{1}^{2} + 3 \, \pi^{2} g_{2}^{2}\right)} g_{4} + \frac{2}{3} \, {\left(32 \, \pi^{2} g_{1}^{2} + 96 \, \pi^{2} g_{2}^{2} - 3 \, \varepsilon\right)} g_{7},\\
        \beta_8&=-\frac{8}{9} \, \pi^{4} g_{1}^{3} + 4 \, \pi g_{2}^{2} + \frac{8}{27} \, {\left(3 \, \pi - 52 \, \pi^{4} g_{1}\right)} g_{5}^{2} + \frac{16}{9} \, {\left(3 \, \pi - 32 \, \pi^{2} g_{1}\right)} g_{5} g_{6} \\
        &+ \frac{8}{9} \, {\left(\pi - 48 \, \pi^{2} g_{1}\right)} g_{6}^{2} + 8 \, {\left(\pi^{2} g_{1}^{2} + 3 \, \pi^{2} g_{2}^{2}\right)} g_{3} + \frac{64}{3} \, {\left(\pi^{2} g_{1}^{2} + 3 \, \pi^{2} g_{2}^{2}\right)} g_{7} + \frac{2}{3} \, {\left(64 \, \pi^{2} g_{1}^{2} + 192 \, \pi^{2} g_{2}^{2} - 3 \, \varepsilon\right)} g_{8}.
    \end{split}
\end{equation}
When we restrict ourselves to $U(N)^3-$invariant interactions (setting $g_1=g_5=g_6=0$), we recover, as expected, the $\beta$ functions from \cite{Benedetti_2020}. 

We now focus on the fixed points defined by the equations
$$\beta_i(g)=0.$$

For vanishing $\varepsilon$, we find a $4$-dimensional manifold of fixed points, spanned by $$(0,0,g_3, g_4, 0,0, g_7, g_8).$$
This manifold of fixed points is the one already found in  \cite{Benedetti_2020}
for the case of the $U(N)$ invariant model. 
As already noticed in \cite{Benedetti_2020}, this manifold  is a generalization of the vector model case, where the interaction is exactly marginal al large $N$.


Let us now analyze the case $\varepsilon>0$. 
We do our fixed point analysis as a function of the behavior of the 
prismatic coupling and of the wheel coupling.

One can distinguish three possibilities:
\begin{enumerate}
    \item The prismatic coupling dominates and the wheel coupling vanishes,
    \item The prismatic coupling vanishes, and the wheel coupling dominates,
    \item Both the prismatic and the wheel coupling vanish.
\end{enumerate}
In the 1st case above, we find a unique Wilson-Fisher fixed point, situated at
\begin{equation}
    \begin{split}
    &g_1^*=\frac{3}{2\pi}\left(\varepsilon-3\varepsilon^2\right),
    ~g_2^*=0,~
    g_3^*={\frac{9}{\pi}} \varepsilon + 2\varepsilon^{2},
    g_4^*=-\frac{3\pi}{4}\varepsilon^{2},~
    g_5^*=0,~
    g_6^*=0,\\
    &g_7^*={\frac{9}{2\pi}} \varepsilon + {\left(-\frac{9}{2}\pi + \frac{747}{2 \pi}\right)} \varepsilon^{2},~
    g_8^*={\left(-\frac{3}{2}\pi + \frac{189}{\pi}\right)} \varepsilon^{2}.
    \end{split}
\end{equation}
This fixed point is the one found in \cite{jepsen2023rg}. Moreover, in \cite{Fraser-Taliente:2024rql}, when the Yukawa interaction is turned off, the authors also find this fixed point, called the \textit{prismatic fixed point}. At this fixed-point the stability matrix $\frac{\partial\beta_i}{\partial g_j}$ can be diagonalized. The eigenvalues at the prismatic fixed-point are:
\begin{equation}
 \Vec{\nu}=   (6\varepsilon, 2\varepsilon,2\varepsilon,-2\varepsilon,-2\varepsilon,-2\varepsilon,-2\varepsilon,-2\varepsilon).
\end{equation}
This indicates that the prismatic fixed point is a saddle point of the RG flow, and this matches the analysis done in \cite{Fraser-Taliente:2024rql}, as well as the eigenvalue 
found 
for the prismatic model in \cite{Giombi_2018}. 

The eigendirections are 
\begin{equation}
\begin{split}
       & (0,0,0,0,0,1,0,0)~,~(0,0,0,0,1,-1,0,0)~,~(0,1,-9,0,0,0,9,-3)~,
       (1,0,6,0,0,0,-1,0)~,\\
       &(0,0,1,0,0,0,-2,1)~,~(0,0,0,1,0,0,-3,2)~,~(0,0,0,0,0,0,1,-1)~,~(0,0,0,0,0,0,0,1).
\end{split}
\end{equation}

In the 2nd case, one finds a pair of fixed-points which already appears 
in the analysis of the
$U(N)^3-$invariant model, with $g_2>0$. These two fixed points are situated at:
\begin{equation}
    \begin{split}
    &g_1^*=0,~
    g_2^*= \pm\frac{\sqrt{\varepsilon}}{2 \, \pi},~
    g_3^*=\frac{9 \, {\left(\mp 2 \, \sqrt{\varepsilon} - 1\right)}}{4 \, \pi},~
    g_4^*=0,\\
   & g_5^*=0,~
    g_6^*=0,~
    g_7^*=\frac{9 \, {\left(\pm7 \, \sqrt{\varepsilon} + 5\right)}}{14 \, \pi},~
    g_8^*= \frac{\mp126 \, \sqrt{\varepsilon} -109}{84 \, \pi}.
    \end{split}
\end{equation}
Both of these fixed-points have the same eigenvalues
\begin{equation}
    \Vec{\nu}=\left(2 \, \varepsilon, 4 \, \varepsilon, 4 \, \varepsilon, 6 \, \varepsilon, 10 \, \varepsilon, 14 \, \varepsilon, 30 \, \varepsilon, 0\right).
\end{equation}

These fixed points are also found in \cite{jepsen2023rg} and \cite{Fraser-Taliente:2024rql} when the Yukawa interaction is turned off. They are called the \textit{wheelic fixed points}. 
Note that the stability matrix is non-diagonalizable at these wheelic fixed points. As indicated in \cite{Benedetti_2020}, this corresponds to a logarithmic CFT (see \cite{Cardy_2013, Creutzig_2013, Hogervorst_2017}), and is therefore non-unitary. 

In the 3rd case, one find a unique trivial fixed point $g_i=0$.
We summarize the short-range fixed points in Table~\ref{FPSR}. \\

\begin{table}[htbp]
\centering
\begin{tabular}{|c|c|c|c|}
\hline
 & $g_1>0$ & $g_2>0$ & $g_1=g_2=0$ \\
\hline
$\varepsilon>0$ & One prismatic fixed point& A pair of wheelic fixed points & Trivial fixed point \\
\hline
$\varepsilon=0$ & No fixed point & No fixed point &$4d$ manifold of fixed points  \\
\hline
\end{tabular}
\caption{Summary of the fixed point structure in the short-range.}
\label{FPSR}
\end{table}

\subsection{Long-range}

Let us now analyze in this subsection the long-range behavior of our model.
We compare here our results with the ones of $U(N)^3$-invariant model. As already mentioned above, let us emphasize here that this long-range behaviour is not analysed in 
\cite{jepsen2023rg} and \cite{Fraser-Taliente:2024rql}.

In the long-range case, one has $\zeta=\frac{d+\varepsilon}{3}$.
The Feynman integrals are (see again \cite{Harribey_2022} for details):
\begin{equation}
    M_3^{\zeta=\frac{d+\varepsilon}{3}}(\mu)=\mu^{-2\varepsilon}\frac{\Gamma(\frac{d}{6})^3}{(4\pi)^d\Gamma(\frac{d}{3})^3\Gamma(\frac{d}{2})}\left(\frac{1}{\varepsilon}+\psi(1)+\psi\left(\frac{d}{6}\right)-2\psi\left(\frac{d}{3}\right)+K)\right)
\end{equation}
with 
\begin{equation}
\begin{split}
\label{K}
        K&=\frac{\Gamma(-d/6)\Gamma(d/3)^2\Gamma(d/2)}{\Gamma(d/6)^2\Gamma(2d/3)}{}_3F_2\!\left(\frac{d}{6},\frac{d}{3},\frac{d}{2};1+\frac{d}{6},\frac{1}{2}+\frac{d}{3};\frac{1}{4} \right)\\
        &+\frac{d^2}{2(d+3)(6-d)}{}_4F_3\!\left(1,1,1+\frac{d}{6},1+\frac{d}{3};2,2-\frac{d}{6},\frac{3}{2}+\frac{d}{6};\frac{1}{4}\right),
\end{split}
\end{equation}
\begin{equation}
    A_1^{\zeta=\frac{d+\varepsilon}{3}}=\mu^{-4\varepsilon}\frac{1}{(4\pi)^{2d}}\frac{\Gamma(d/6)^9}{\Gamma(d/3)^9\Gamma(d/2)}\left(\frac{1}{2\varepsilon}+\cO(\varepsilon^0)\right),
\end{equation}
\begin{equation}
    A_2^{\zeta=\frac{d+\varepsilon}{3}}=\mu^{-4\varepsilon}\frac{\Gamma(d/6)^6}{2(4\pi)^{2d}\Gamma(d/2)^2\Gamma(d/3)^6}\left[\frac{1}{\varepsilon^2}+\frac{1}{\varepsilon}\left(3\psi(1)+\psi(d/6)-5\psi(d/3)+\psi(d/2)+K\right)\right], 
\end{equation}
\begin{equation}
    A_3^{\zeta=\frac{d+\varepsilon}{3}}=\Big(M_3^{\zeta=\frac{d+\varepsilon}{3}}(\mu)\Big)^2,
\end{equation}
\begin{equation}
    A_4^{\zeta=\frac{d+\varepsilon}{3}}=-\mu^{-4\varepsilon}\frac{2\pi^2}{\varepsilon(4\pi)^6}+\cO(\varepsilon^0),
\end{equation}
where in \eqref{K}, ${}_pF_q$ is a generalized hypergeometric function.

Inserting these integrals in eq.~\eqref{betageneric} gives, after some tedious but straightforward algebra, the explicit expressions of  the long-range $\beta$ functions. Note that in the long-range, we only study the $\varepsilon=0$ case, analogously to \cite{Benedetti_2020}.
After the rescalings
\begin{equation}
    \tilde{g}=g(4\pi)^d, 
\end{equation}
and 
\begin{equation}
    \tilde{\beta}=\frac{\beta}{(4\pi)^d},
\end{equation}
the long-range $\beta$ functions read:
\begin{equation}
    \begin{split}
        \beta_1&=\frac{2g_1^2}{3}\alpha_1,\\
        \beta_2&=0,\\
        \beta_3&=\frac{g_1^3}{3}(8\alpha_2+4\alpha_3+8\alpha_4)+4\left(g_1^2+\frac{g_5^2}{9}+\frac{9g_2^2}{2} \right)\alpha_1-4\left(3g_1^2g_2+3g_1g_2^2+3g_1^3+9g_2^3+\frac{g_1^2g_3}{3}+g_2^2g_3\right)\alpha_5,\\
        \beta_4&=-2(g_1^2+3g_2^2)g_4\alpha_5+\frac{2g_5^2}{9}\alpha_1-\frac{2g_1^3}{9}\alpha_6-\frac{2g_1g_5^2}{27}(\alpha_3-2\alpha_2-2\alpha_4),\\
        \beta_5&=\frac{4g_1g_5}{3}\alpha_1-\frac{2}{9}(g_1^2+3g_2^2)g_5\alpha_5,\\
        \beta_6&=\frac{4g_1}{3}(g_5+2g_6)\alpha_1-2g_2^2(4g_5+5g_6)\alpha_5,\\
        \beta_7&=(2g_1^3+\frac{14}{9}g_5^2+\frac{8}{3}g_5g_6)\alpha_1-\frac{4g_1^3}{3}\alpha_6-\frac{2}{3}(g_1^2+3g_2^2)(27g_2+10g_3+12g_4+7g_7+9g_1)\alpha_5\\
        &+\left(\frac{104}{27}g_1g_5^2+\frac{4}{3}g_1^3+\frac{32}{9}g_1g_5g_6\right)\left(\alpha_2-\frac{\alpha_3}{2}+\alpha_4\right),\\
        \beta_8&=\left(2g_2^2+\frac{4g_5^2}{9}+\frac{8}{3}g_5g_6+\frac{4}{9}g_6^2\right)\alpha_1-\frac{4g_1^3}{9}\alpha_6-\frac{2}{3}(g_1^2+3g_2^2)(3g_3+8g_7+15g_8)\alpha_5\\
        &+\left(\frac{16}{3}g_1g_6^2+\frac{64}{9}g_1g_5g_6\right)\left(\alpha_2-2\alpha_3+\alpha_4\right), 
    \end{split}
\end{equation}
where 
\begin{equation}
\begin{split}
    \alpha_1&=\frac{\Gamma(\frac{d}{6})^3}{\Gamma(\frac{d}{2})\Gamma(\frac{d}{3})},\\
    \alpha_2&=\frac{\Gamma(\frac{d}{6})^4\Gamma(-\frac{d}{6}){}_3F_2\!\left(\frac{d}{6},\frac{d}{3},\frac{d}{2};1+\frac{d}{6},\frac{1}{2}+\frac{d}{3};\frac{1}{4} \right)}{\Gamma(\frac{2d}{3})\Gamma(\frac{d}{2})\Gamma(\frac{d}{3})^3},\\
    \alpha_3&=\frac{d^2\Gamma(\frac{d}{6})^6{}_4F_3\!\left(1,1,1+\frac{d}{6},1+\frac{d}{3};2,2-\frac{d}{6},\frac{3}{2}+\frac{d}{6};\frac{1}{4}\right)}{(d+3)(d-6)\Gamma(\frac{d}{2})^2\Gamma(\frac{d}{3})^6},\\
    \alpha_4&=\frac{\Gamma(\frac{d}{6})}{\Gamma(\frac{d}{2})^2\Gamma(\frac{d}{6})^6}\left(\psi\left(\frac{d}{6}\right)-\psi(1)+\psi\left(\frac{d}{3}\right)-\psi\left(\frac{d}{2}\right) \right),\\
    \alpha_5&=\frac{\Gamma(\frac{d}{6})^4\Gamma(-\frac{d}{6})}{\Gamma(\frac{2d}{3})\Gamma(\frac{d}{2})\Gamma(\frac{d}{3})^4},\\
    \alpha_6&=\frac{\Gamma(\frac{d}{6})^9}{\Gamma(\frac{d}{2})\Gamma(\frac{d}{3})^9}.
\end{split}
\end{equation}
Let us mention that, analogously to the $U(N)^3$ case, the wheel $\beta$ function vanishes in this case.

We now focus on the analysis of fixed points. First, we find the $4-$dimensional manifold of fixed points
spanned by 
$$(0,0,g_3,g_4,0,0, g_7,g_8),$$ 
where both the prismatic and the wheel coupling are zero, as 
 found in the short-range case (see the previous subsection).
We also find the line of fixed-points, parameterized by the wheel coupling $g_2$:
\begin{equation}
    \begin{split}
    &g_1^*=0,~
    g_2^*=x,~
    g_3^*=-9x+\frac{9\Gamma(2d/3)\Gamma(d/3)}{2\Gamma(d/6)\Gamma(-d/6)},~
    g_4^*=0,~
    g_5^*=0,~\\
    &g_6^*=0,~
    g_7^*=9x-\frac{45\Gamma(2d/3)\Gamma(d/3)}{7\Gamma(d/6)\Gamma(-d/6)},~
    g_8^*=-3x+\frac{327\Gamma(2d/3)\Gamma(d/3)}{\Gamma(d/6)\Gamma(-d/6)}.
    \end{split}
\end{equation}

Note that 
the fixed points 
found here 
are the same as the ones found in the case of $U(N)^3-$invariant sextic tensor model.  The stability matrix at these fixed points can also be diagonalized, so that the critical exponents are 
\begin{equation}
    \Vec{\nu}=\cC\left( 30,14,10,6,4,2,0,0\right), 
\end{equation}
with
\begin{equation}
    \cC=-x^2\frac{\Gamma(-d/6)\Gamma(d/6)^4}{\Gamma(2d/3)\Gamma(d/2)\Gamma(d/3)^4}.
\end{equation}
The eigendirections are
\begin{equation}
\begin{split}
\label{eigdir}
    &(0,0,0,0,0,0,0,1)~,~(0,0,0,0,0,0,1,-1)~,~(0,0,0,0,0,1,0,0)~,~(0,0,0,1,0,0,-3,2)~\\
    &(0,0,1,0,0,0,-2,1)~,~(0,0,0,0,1,-1,0,0)~,~(1,0,-3,0,0,0,3,-1)~,~(0,1,-9,0,0,0,9,-3).
\end{split}
\end{equation}
Again, and as in the $U(N)^3$ case, the critical exponents are positive, indicating that the scaling operators defined in the directions of \eqref{eigdir} are irrelevant. 
No other fixed points are found.

We conclude that the long-range structure 
of the fixed points of the $O(N)^3$ model is 
therefore 
identical to the one found 
for the $U(N)^3$ model. This thus means that the prismatic interaction, as well as the other supplementary interactions of our model,
do not generate new fixed points in the long-range case. 

\section{Conclusion}
\label{sec5}

In this paper, we have studied the RG flows of the $O(N)^3$-invariant general sextic tensor field theory  with optimal scalings of the $8$ interacting terms of the action. We computed the $\beta$ functions at cubic order in the coupling constant and at leading order in the large $N$ limit, employing a standard diagrammatic expansion in dominant $6$-point graphs rather than the multi-scalar approach previously used in the literature. Our analysis applies to both the short-range ($\zeta=1$) and long-range ($\zeta<1$) regimes. 
In the short-range case, we recovered, as expected, the three fixed points identified in \cite{jepsen2023rg, Fraser-Taliente:2024rql}: the two wheelic fixed points and the prismatic fixed point. 

In the long-range case, we found that the fixed-point structure is characterized by a continuous line of fixed points, which is identical 
to the one found in
the $U(N)^3$ model studied in \cite{Benedetti_2020}. 

From the point of view of Feynman graph diagrammatics, 
this result can appear as surprising, 
since the leading order structure of the $O(N)^3$ model is 
significantly richer than the one of  $U(N)^3$ model (see the analysis of 
\cite{largeN}). Thus, in the $O(N)^3$ case, the dominant graph structure includes contributions from melonic graphs built on non-bipartite bubbles, such as those depicted in Figs.~\ref{fig:I5fromM3} and~\ref{fig:I6fromM3}, which generate intricate index contraction patterns absent in the bipartite case. 

However, the results obtained in this paper may appear less surprising when the fixed-point structure is viewed as a mean of identifying the CFTs compatible with the dynamics of the theory under consideration. This question can, in fact, be addressed in a much broader setting through the celebrated $F$-theorem.

Let us recall here that the $F$-theorem states that the sphere free energy of a field theory decreases between the UV and IR fixed points.
In the context of tensor field theories such as the one studied in this paper, several interesting results have been obtained in \cite{F} and \cite{W}.

Thus, in \cite{F}, the long-range bosonic $O(N)^3$-invariant tensor model on a spherical background was considered at NNLO in the $1/N$ expansion. The $F$-theorem was tested and confirmation that the theorem holds in this case was obtained.

Moreover, in \cite{W}, it was shown that the conformal data of a range of large-$N$ CFTs (range which includes generalized SYK models, vector models and tensor field theories) are specified by constrained extremization of the universal part of the sphere free energy. This result was established using the 2PI effective action and, equivalently, by Feynman diagram resummation.

\medskip

Let us now end this paper by listing a potential 
direction for future investigations. It appears 
interesting to us 
to understand whether or not there exists a connection between the dominant graph structure in the $1/N$ expansion and the fixed-point structure of the theory. The question to be asked in this context is: can one infer properties of fixed points from the combinatorial structure of dominant graphs? 


\appendix

\section{Loop integrals}
\label{Appint}
\paragraph{Self energy integral}
The following integral is useful in various computations\footnote{We use the notation:  $\int_k=\int\frac{\text{d}^dk}{(2\pi)^d}$.}:
\begin{equation}
    \int_k\frac{1}{k^{2\alpha}(k+p)^{2\beta}}=\frac{p^{-2(\alpha+\beta-\frac{d}{2})}}{(4\pi)^{\frac{d}{2}}}\frac{\Gamma(\frac{d}{2}-\alpha)\Gamma(\frac{d}{2}-\beta)\Gamma(\alpha+\beta-\frac{d}{2})}{\Gamma(\alpha)\Gamma(\beta)\Gamma(d-\alpha-\beta)}.
   \label{master}
\end{equation}

\paragraph{General melonic integral}
We can define the r-melonic integral  $M_r^\zeta(p)$, that carries momentum p with Laplacian power $\zeta$:
\begin{equation}
    M^\zeta_r(p)=\int\prod_{i=1}^r\left(\frac{\text{d}^dq_i}{(2\pi)^d}\frac{1}{q_i^{2\zeta}}\right)\frac{1}{|\sum_{i=1}^r q_i+p|^{2\zeta}}.
\end{equation}

We show by induction that: 
\begin{equation}
    M_r^{\zeta}(p)=\frac{p^{(r-1)d-2r\zeta}}{(4\pi)^\frac{(r-1)d}{2}}\frac{\Gamma(\frac{d}{2}-\zeta)^r}{\Gamma(\zeta)^r}\frac{\Gamma(\frac{1}{2}(2r\zeta-(r-1)d))}{\Gamma(\frac{r}{2}(d-2\zeta))}\quad,\forall r\geq2.
    \label{generalmelon}
\end{equation}
We start by $M_2^\zeta(p)$, using eq.~\eqref{master} to integrate explicitly. 
\begin{equation}
    M_2^\zeta(p)=\frac{p^{d-4\zeta}}{(4\pi)^\frac{d}{2}}\frac{\Gamma(\frac{d}{2}-\zeta)^2}{\Gamma(\zeta)^2}\frac{\Gamma(2\zeta-\frac{d}{2})}{\Gamma(d-2\zeta)},
\end{equation}
which matches eq.~\eqref{generalmelon} for $r=2$. One then has
\begin{equation}
\begin{split}
       M^\zeta_{r+1}(p)&=\int\prod_{i=1}^{r}\left(\frac{\text{d}^dq_i}{(2\pi)^d}\frac{1}{q_i^{2\zeta}}\right)\frac{1}{|\sum_{i=1}^{r} q_i+p|^{2\zeta}}\\
       &=\int\frac{\text{d}^dq_1}{(2\pi)^d}\frac{1}{q_1^{2\zeta}}M_r^\zeta(q_1+p)\\
       &=\frac{1}{(4\pi)^\frac{(r-1)d}{2}}\frac{\Gamma(\frac{d}{2}-\zeta)^r}{\Gamma(\zeta)^r}\frac{\Gamma(\frac{1}{2}(2r\zeta-(r-1)d))}{\Gamma(\frac{r}{2}(d-2\zeta))}\int_{q_1}\frac{1}{q_1^{2\zeta}}\frac{1}{(q_1+p)^{2(r\zeta-\frac{r-1}{2}d)}}\\
       &=\frac{p^{rd-2(r+1)\zeta}}{(4\pi)^{\frac{(r+1)d}{2}}}\frac{\Gamma(\frac{d}{2}-\zeta)^{r+1}}{\Gamma(\zeta)^{r+1}}\frac{\Gamma(\frac{1}{2}(2(r+1)\zeta-rd))}{\Gamma(\frac{r+1}{2}(d-2\zeta))},
\end{split}
\end{equation}
where we used eq.~\eqref{master} to integrate in the third line. This matches eq.~\eqref{generalmelon} at rank $r+1$.

\begin{figure}[H]
    \centering
    \includegraphics[width=0.4\textwidth]{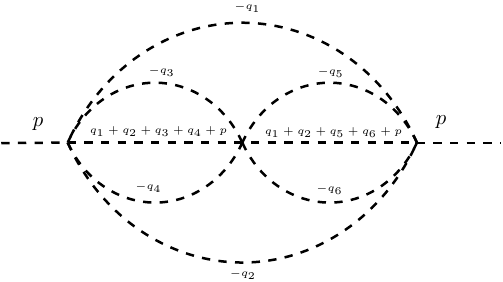}
    \caption{$2$-point glasses diagram with internal momenta. This choice of momenta allows us to factorize $\big(M_3^\zeta(q_1+q_2+p)\big)^2$.}
    \label{fig:2PglassesMomenta}
\end{figure}

\paragraph{Glasses integral} 
We compute the glasses integral $\text{G}^\zeta(p)$ for the diagram depicted in Fig. \ref{fig:2PglassesMomenta}.

The integral is
\begin{equation*}
\begin{split}
        \text{G}^\zeta(p)&=\int_{q_1q_2}G(q_1)G(q_2)\Big(\int_{q_3q_4}G(q_3)G(q_4)G(q_1+q_2+q_3+q_4+p)\Big)\Big(\int_{q_5q_6}G(q_5)G(q_6)G(q_1+q_2+q_5+q_6+p)\Big)\\
        &=\int_{q_1,q_2}G(q_1)G(q_2)(M_3^\zeta(q_1+q_2+p))^2.
\end{split}
\end{equation*}

Using eq.~\eqref{generalmelon} for $r=3$, one gets:
\begin{equation}
    M_3^\zeta(q_1+q_2+p)=\frac{1}{(4\pi)^d}\frac{\Gamma(\frac{d}{2}-\zeta)^3}{\Gamma(\zeta)^3}\frac{\Gamma(3\zeta-d)}{\Gamma(\frac{3d}{2}-3\zeta)}\frac{1}{|p+q_1+q_2|^{2(3\zeta-d)}}.
\end{equation}
Plugging in $\text{G}^\zeta(p)$ and again integrating loop by loop using eq.~\eqref{master}, the integral becomes:
\begin{equation}
    \begin{split}
        \text{G}^\zeta(p)=&\frac{1}{(4\pi)^{2d}}\frac{\Gamma(\frac{d}{2}-\zeta)^6}{\Gamma(\zeta)^6}\frac{\Gamma(3\zeta-d)^2}{\Gamma(\frac{3d}{2}-3\zeta)^2}\int_{q_1,q_2}\frac{1}{q_1^{2\zeta}}\frac{1}{q_2^{2\zeta}}\frac{1}{|p+q_1+q_2|^{2(6\zeta-2d)}}\\
        &=\frac{1}{(4\pi)^{3d}}\frac{\Gamma(\frac{d}{2}-\zeta)^8}{\Gamma(\zeta)^8}\frac{\Gamma(3\zeta-d)^2}{\Gamma(\frac{3d}{2}-3\zeta)^2}\frac{\Gamma(\frac{5d}{2}-6\zeta)}{\Gamma(6\zeta-2d)}\frac{\Gamma(8\zeta-3d)}{\Gamma(\frac{7d}{2}-8\zeta)}\frac{1}{p^{2(8\zeta-3d)}}.
    \end{split}
\end{equation}

\section{$\beta$ functions}
\label{AppBeta}
The $\beta$ functions are obtained through a diagrammatic expansion of the $6$-point vertex functions, as described in the body text. 
One gets:
\begin{equation}
    \begin{split}
    \label{betageneric}
        \beta_1&=(-2\varepsilon+3\eta)g_1-\frac{1}{6}\left(\mu^{2\varepsilon}g_1^2+\frac{2\mu^{4\varepsilon}g_1^3}{6}M_3(\mu)\right)\dlog M_3(\mu)+\frac{\mu^{4\varepsilon}g_1^3}{36}\dlog A_3(\mu),\\
        \beta_2&=(-2\varepsilon+3\eta)g_2,\\
        \beta_3&=(-2\varepsilon+3\eta)g_3-2\left(\mu^{2\varepsilon}g_1^2+\frac{\mu^{4\varepsilon}g_1^3}{3}M_3(\mu)\right)\dlog M_3(\mu)+9\mu^{2\varepsilon}g_2^2\dlog M_3(\mu)\\
        +&\frac{2\mu^{4\varepsilon}g_1^3}{3}\dlog A_2(\mu)
        +\mu^{4\varepsilon}\left(2g_1^3+6g_1^2g_2+6g_1g_2^2+18g_2^3+\frac{g_1^2g_3}{3}+2g_2^2g_3\right)\dlog A_4(\mu),\\
        \beta_4&=(-2\varepsilon+3\eta)g_4-\frac{1}{9}\left(\mu^{2\varepsilon}g_5^2+\frac{4\mu^{4\varepsilon}g_1g_5^2}{3}M_3(\mu)\right)\dlog M_3(\mu)+\frac{\mu^{4\varepsilon}g_1^3}{9}\dlog A_1(\mu)\\
        &+\mu^{4\varepsilon}(g_1^2+3g_2^2)g_4\dlog A_4(\mu)+\frac{\mu^{4\varepsilon}g_1g_5^2}{27}\left(2\dlog A_2(\mu)+\dlog A_3(\mu)\right),\\
        \beta_5&=(-2\varepsilon+3\eta)g_5-\frac{2}{3}\left(\mu^{2\varepsilon}g_1g_5+\mu^{4\varepsilon}g_1^2g_5M_3(\mu)\right)\dlog M_3(\mu)+\frac{\mu^{4\varepsilon}g_1^2g_5}{3}\dlog A_3(\mu)\\
        &+\mu^{4\varepsilon}\left(\frac{g _1^2g_5}{9}+g_2^2g_5\right)\dlog A_4(\mu),
        \end{split}
    \end{equation}
    \begin{equation*}
    \begin{split}
        \beta_6&=(-2\varepsilon+3\eta)g_6-\frac{4}{3}\left(\mu^{2\varepsilon}g_1g_6+\mu^{4\varepsilon}g_1^2\left(2g_6+\frac{2g_5}{3}\right)M_3(\mu)\right)\dlog M_3(\mu)\\
        &-\frac{2}{3}(\mu^{2\varepsilon}g_1g_5+\mu^{4\varepsilon}g_1^2g_5M_3(\mu))\dlog M_3(\mu)+\mu^{4\varepsilon}g_1^2\left(\frac{3g_6}{4}+\frac{g_5}{2}\right)\dlog A_3(\mu)\\
        &+\mu^{4\varepsilon}\left(\frac{4g_1^2g_5}{3}+\frac{5g_1^2g_6}{3}+4g_2^2g_5+5g_2^2g_6\right)\dlog A_4(\mu),\\
        \beta_7&=(-2\varepsilon+3\eta)g_7-\left(\mu^{2\varepsilon}g_1^2+\frac{2\mu^{4\varepsilon}g_1^3}{6}M_3(\mu)\right)\dlog M_3(\mu)-\frac{7}{9}\Bigg(\mu^{2\varepsilon}g_5^2\\
        &+\frac{4\mu^{4\varepsilon}g_1g_5^2}{3}M_3(\mu)\Bigg)\dlog M_3(\mu) -\frac{4}{3}\left(\mu^{2\varepsilon}g_5g_6+\frac{8\mu^{4\varepsilon}g_1g_5g_6}{3}M_3(\mu)+\mus g_1g_5^2M_3(\mu)\right)\dlog M_3(\mu)\\
        &+\muf\Bigg(\frac{g_1^3}{3}(2\dlog A_1(\mu)+\dlog A_2(\mu))+\Big(3g_1^3+9g_1^2g_2+\frac{10g_1^2g_3}{3}+4g_1^2g_4+9g_2^2g_1+27g_2^3+10g_2^2g_3\\
        &+12g_2^2g_4+\frac{7g_1^2g_7}{3}+7g_2^2g_7\Big)\dlog A_4(\mu)+\frac{16g_1g_5g_6}{9}(\dlog A_2(\mu)+2\dlog A_3(\mu))+\frac{52g_1g_5^2}{27}\dlog A_2(\mu)\Bigg), \\
        \beta_8&=(-2\varepsilon+3\eta)g_8+\Bigg( -g_2^2\mu^{2\varepsilon} -\frac{2}{9}(g_6^2\mu^{2\varepsilon}+2\muf g_1^2g_6M_3(\mu))-\frac{4}{3}(\mu^{2\varepsilon} g_5g_6\\
        &+\frac{8\muf g_1g_5g_6}{3}M_3(\mu)+\frac{2\muf g_1g_5^2}{3}  M_3(\mu))-\frac{2}{9}\left(\mu^{2\varepsilon} g_5^2+\frac{4\muf g_1g_5^2}{3}M_3(\mu)\right)
        \Bigg)\dlog M_3(\mu) \\
        &+\muf\Big(\frac{2g_1^3}{9}\dlog A_1(\mu)+\big(\frac{220g_1g_5g_6}{81}+\frac{8g_1g_6^2}{3}\big)\dlog A_2(\mu)+\frac{10}{27}g_1(g_6^2+2g_5^2)\dlog A_3(\mu)\\
        &+\big(g_1^2g_3+\frac{8g_1^2g_7}{3}+5g_1^2g_8+3g_2^2g_3+8g_2^2g_7+15g_2^2g_8\big)\dlog A_4(\mu)\Big).
    \end{split}
\end{equation*}

\section{The $\beta_4$ computation}
\label{AppBeta4}In this section we explicitly compute the function $\beta_4$ in order to illustrate the general derivation procedure outlined in the body text.

\begin{figure}
    \centering
    \includegraphics[scale=0.7]{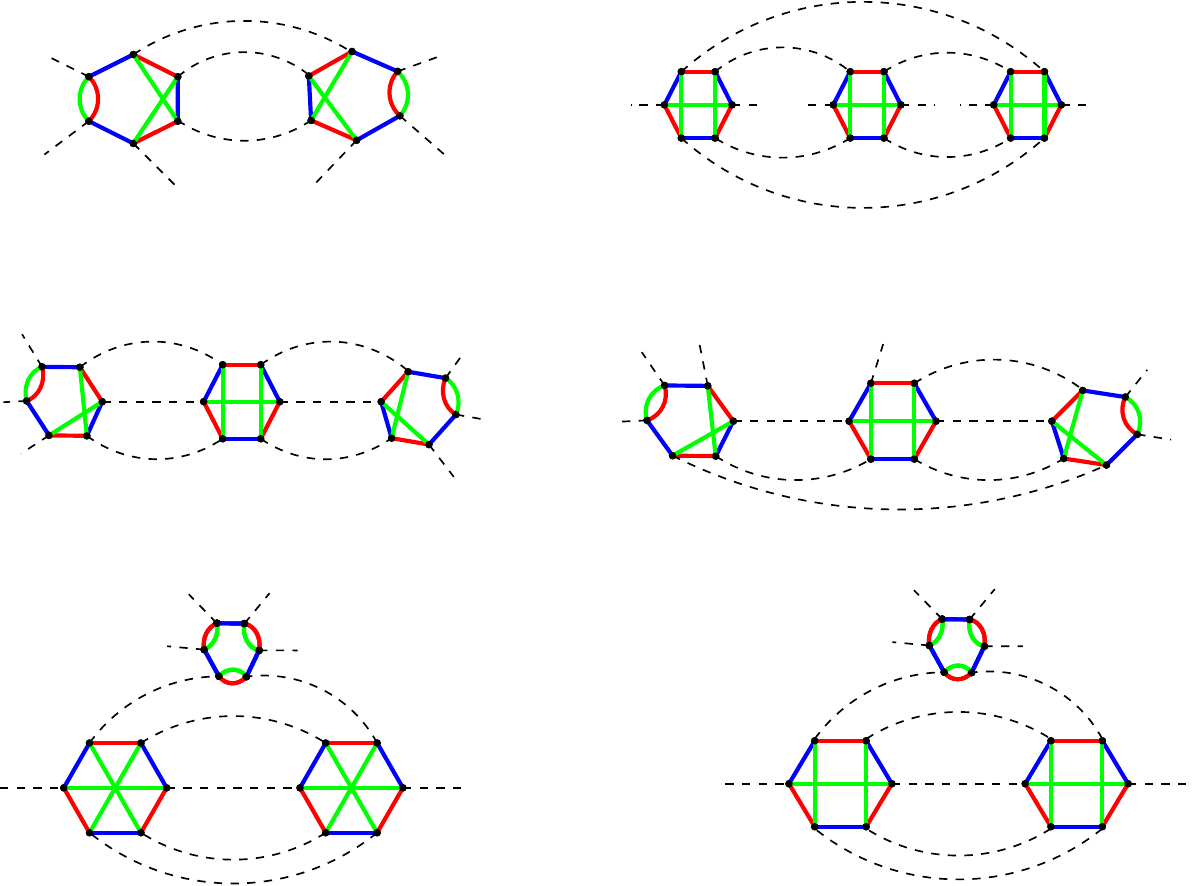}
    \caption{The six dominant $6$-point graphs with $I_4$ external index structure contributing at order $\mathcal{O}(\lambda^3)$ in the large-$N$ limit.}
    \label{fig:gamma4}
\end{figure}

The first step is to identify the diagrammatic expansion of the $6$-point vertex function $\Gamma_4^{(6)}$ with external index structure $I_4$. At leading order in $1/N$ and up to order $\mathcal{O}(\lambda^3)$, there are six dominant graphs contributing to this vertex function. These graphs are depicted in Fig.~\ref{fig:gamma4}. One can verify explicitly that the degree of each of these graphs is
\begin{equation}
    \omega=7=3+\alpha_4,
\end{equation}
confirming that they scale identically to the $I_4$ interaction and are therefore dominant contributions in the large $N$ limit.

The $6$-point vertex function for the $I_4$ interaction is given by
\begin{equation}
    \Gamma_4^{(6)}=\lambda_4-\frac{\lambda_5^2}{9}M_3(\mu)+\frac{\lambda_1^3}{9}A_1(\mu)+(\lambda_1^2+3\lambda_2^2)\lambda_4 A_4(\mu)+\lambda_1\lambda_5^2\frac{2A_2(\mu)+A_3(\mu)}{27},
\end{equation}
where $M_3(\mu)$ denotes the melon integral and $A_i(\mu)$ are the loop integrals defined in the main text.

Recalling the definition of the renormalized coupling,
\begin{equation}
    g_4=\mu^{-2\varepsilon} Z^3 \Gamma_4^{(6)}(\{p_j\}, \mu, \{\lambda\}), 
\end{equation}
we obtain the $\beta$ function by differentiating with respect to the renormalization scale:
\begin{equation}
\begin{split}
        \beta_4&=\mu\partial_\mu g_4=-2\varepsilon g_4+\frac{3}{Z}g_4\partial_\mu Z +\mu^{-2\varepsilon}Z^3 \mu\partial_\mu \Gamma_4^{(6)}\\
        &=(-2\varepsilon+3\eta)g_4+\mu^{-2\varepsilon}\Bigg(-\frac{\lambda_5^2}{9}\mu\partial_\mu M_3(\mu)+\frac{\lambda_1^3}{9}\mu\partial_\mu A_1(\mu)\\
        &\qquad+(\lambda_1^2+3\lambda_2^2)\lambda_4 \mu\partial_\mu A_4(\mu)+\lambda_1\lambda_5^2\frac{2\mu\partial_\mu A_2(\mu)+\mu\partial_\mu A_3(\mu)}{27} \Bigg).
        \label{beta4_inter}
\end{split}
\end{equation}
Above we have used $Z^3=1+\cO(\lambda^3)$ and the fact that $\partial_\mu \lambda_i=0$ for the bare couplings.

The expression in eq.~\eqref{beta4_inter} still contains the bare couplings $\lambda_1$, $\lambda_2$ and $\lambda_5$. To express the $\beta$ function in terms of renormalized couplings $g_i$, we must substitute the inverted bare expansion $\{\lambda_i(g)\}$ into eq.~\eqref{beta4_inter}. 
For the couplings $\lambda_1$ and $\lambda_2$, the inverted bare expansions are given in eqs.~\eqref{invg1} and~\eqref{invg1g2}, respectively. 

\begin{figure}[H]
    \centering
    \includegraphics{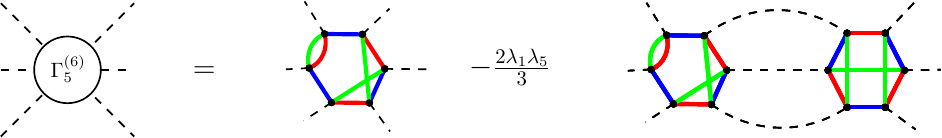}
    \caption{Perturbative expansion of the $6-$point function $\Gamma_5^{(6)}$, up to quadratic order in $\lambda$. }
    \label{fig:gamma5}
\end{figure}

To obtain the inverted expansion for $\lambda_5$, one needs to compute the $6-$point function $\Gamma_5^{(6)}$. This $6-$point function for the $I_5$ interaction is represented in Fig.~\ref{fig:gamma5}. 

Eq.~\eqref{renormalized couplings} for the $g_5$ coupling therefore reads
\begin{equation}
    g_5 = \mu^{-2\varepsilon}Z^3\left(\lambda_5-\frac{2}{3}\lambda_1\lambda_5 M_3 + \cO(\lambda^3)\right).
\end{equation}
This can be perturbatively inverted, yielding
\begin{equation}
    \lambda_5 = \mu^{2\varepsilon}g_5+\frac{2}{3}\mu^{4\varepsilon}g_1g_5 M_3.
\end{equation}

We can therefore writes:
\begin{equation}
    \begin{split}
       & \lambda_5^2=\mu^{4\varepsilon}g_5^2+\frac{4}{3}\mu^{6\varepsilon}g_1g_5^2M_3+\cO(g^4),\\
        &\lambda_1^2 =\mu^{4\varepsilon}g_1^2+\frac{1}{3}\mu^{6\varepsilon}g_1^3M_3+\cO(g^4),\\
        &\lambda_2^2=\mu^{4\varepsilon}g_2^2+\cO(g^4),\\
        &\lambda_1^3=\mu^{6\varepsilon}g_1^3+\cO(g^4),\\
        &\lambda_1\lambda_5^2=\mu^{6\varepsilon}g_1g_5^2+\cO(g^4),
    \end{split}
\end{equation}
in eq.~\eqref{beta4_inter}. This gives the final expression of the $\beta$ function in terms of renormalized couplings:
\begin{equation}
\begin{split}
 \beta_4&=(-2\varepsilon+3\eta)g_4-\frac{1}{9}\left(\mu^{2\varepsilon}g_5^2+\frac{4\mu^{4\varepsilon}g_1g_5^2}{3}M_3(\mu)\right)\dlog M_3(\mu)+\frac{\mu^{4\varepsilon}g_1^3}{9}\dlog A_1(\mu)\\
        &+\mu^{4\varepsilon}(g_1^2+3g_2^2)g_4\dlog A_4(\mu)+\frac{\mu^{4\varepsilon}g_1g_5^2}{27}\left(2\dlog A_2(\mu)+\dlog A_3(\mu)\right).
\end{split}
\end{equation}

\bibliographystyle{ieeetr} 
\bibliography{renorm.bib}
\end{document}